# Clues to detect the Pilot Wave in a photon double-slit interference experiment

Andrea Petrucci

ENEA - Italian National Agency for New Technologies, Energy and Sustainable Economic Development, Via Anguillarese, 301 – 00123 Roma, Italy;
GNFM, Istituto Nazionale di Alta Matematica "F.Severi", Città Universitaria, P.le A.Moro 2, 00185 Roma, Italy,
petrucciandr@gmail.com; andrea.petrucci@enea.it

**Abstract:** The Young double-slit interference pattern produced by quantum objects, like photons, that move through a double-slit is regarded, by the conventional Copenhagen interpretation of Quantum Mechanics, as the evidence of the wave-like behaviour potentially contained in the wave function. On the contrary, a more realistic view of this phenomenon considers the quantum object a particle accompanied by a pilot wave which would be the cause of the interference fringes. This paper proposes a feasible experiment, based on an easy variation of the nowadays common double-slit experimental set-ups, aimed at detecting the effects of the pilot wave once "detached" from the particles that it steers. Besides, a further realistic idea, based on the geometrical violation of Local Lorentz Invariance (LLI), is put forward as to the intrinsic nature of the photon. This new idea along with the possibly positive results of the experiment would allow us to shed new light on the real nature of quantum objects in term of the geometrical violation of LLI.

**Introduction**

The Volta and the Solvay congresses both held in 1927, the former in Como (Italy), the latter in Brussels (Belgium) are the places of birth of the Quantum Mechanics as an axiomatic theory based on the concepts of probability, indeterminacy, complementarity, correspondence. As a consequence of this birth, two streams of thought were born. Some physicists agreed with Bohr and preferred to have some kind of purely conventional recipes to make calculations with no connection to the physical phenomenon, which was said to be indeterminate until the actual measurement; some other physicists strongly disagreed with this vision and preferred to believe that reality is independent of its observation and went on searching for a more complete and satisfactory description and explanation of quantum effects. Emblematic of the conflict between these two currents of thought was the wave-corpuscle duality of a quantum object like for instance the photon or the electron. In some experiments they behaved like a particle (Compton effect for the photons, traces in a cloud chamber for the electron), in some others they produced interference fringes on a photographic plate (the double-slit experiments, the Devisson-Germer experiment). Physicists like Einstein and de Broglie, who were in favour of the existence of a real physical phenomenon and did not want to consider the particle and the wave only as conventional concepts borrowed from the classical physics and applied with some indeterminacy to the quantum physics, tried to look at the particle and the wave-like behaviours as the sign that the reality of a quantum object was actually something more complex that included both natures. Under this light de Broglie minted the name of *pilot wave* to indicate that the particle should not be considered alone but that it was accompanied and actually steered in its motion by a wave which was responsible of the wave-like behaviour of the particle. However the experiments showed that it was the particle to carry all the energy and momentum of the quantum object and for this reason this pilot wave was also called, by Einstein, *hollow wave*, that is a wave that did produce some effects but only on the particle that it piloted, not in any other direct measurements of its own effects.

This paper would like to shed some light on the old debated question as to the reality of the pilot wave and hence to the existence of a deeper comprehension of quantum phenomena. This target will be aimed at both and above all by proposing an experiment with photons intended for detecting the effects of hollow waves deprived of their particles and by putting forward some clues as to the interpretation of the pilot wave in terms of a possible geometrical violation of the local Lorentz invariance.

**Double-slit experiments**

Double-slit experiments either with photons or with electrons have been performed loads of times since the discovery of the wave-corpuscle duality of a quantum object. Being deeply connected to the principle of indeterminacy, several attempts were carried out with the purpose to obtain the interference fringes together with the which-way information, that is the information to know from which of the two slits the particle had gone through. With the improvement of the technology, these experiments have become very accurate and precise [1] but have received anyway negative remarks from those who agreed with the conventional view of quantum mechanics. A further improvement of the techniques to emit electrons or photons to be focalised in a beam on the double-slit made it possible to resolve the bright fringes into the single tiny bright spots made by each single particle on its arrival on the photographic plate. This result was obtained by extremely reducing the number of particles in the beam. On the photographic plate appeared now dark and bright fringes, the wave-like behaviour, and single spots in each bright fringe, the corpuscle-like behaviour. Several experiments of this type have been performed so far either with photons [1] or with electrons [2] and they all indicate that the quantum object (either photon or electron or any other) is actually a more complex thing since in the same experimental set-up there appear evidences belonging to the wave, the fringes, and to the particle, the spots. Despite the negative remarks that also these experiments have received again in favour of the conventional view of quantum mechanics, this kind of experiments questions seriously the completeness of information contained in the wave function and its collapse and open new perspectives and possibilities to the realistic view of de Broglie, Einstein and Bohm of the existence of a pilot wave. In the last decades the capabilities of reducing the number of particles in the beam have further increased and allowed to perform these double-slit experiments in the limit of single photon present in the apparatus during the time of flight from the source to the photographic screen. Also in this case one obtains the fringes and the spots and the conclusion is that the photon somehow interfere with itself.

**Pilot Wave detection: experimental set-up**

De Broglie and Andrade y Silva proposed [3] once an experiment in order to determine the effects of the pilot wave. To the best of my knowledge, an adaptation of this experiment was carried out for the first time in l'Aquila in 1999 and was reported in reference [4]. It was then repeated two more times by different people of the same team and with different experimental set-ups [5,6]. The authors connect the effect measured in the experiment to the violation of Local Lorentz Invariance (LLI) that would be the ultimate explanation of the pilot wave, but for now let's disregard this interpretation and focus the attention on the pilot waves of photons. I will briefly return to this connection in the following paragraph. I will not describe the whole experimental set-up, for which one may refer to [4-8], but only mention what is important to modify in the double-slit set-up for photons in order to measure the effect of the pilot wave. Fig.1 shows the experimental apparatus of [4,6]. It is the content of a box that comprises: two Light Emitting Diodes (LEDs) in the infra-red range, the sources; three photodiodes, the detectors; several panels that divide the space into several rooms; three apertures F1, F2 and F3.

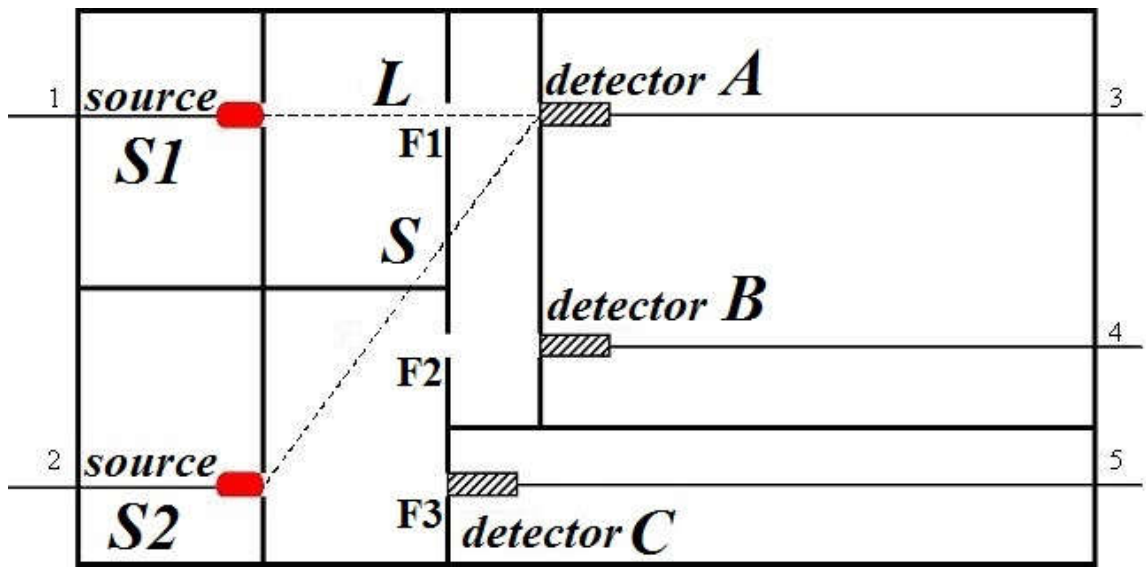

Fig.1. Experimental apparatus to detect the effects of the pilot waves of photons

The detector A collects the photons from S1, the detector C collects the photons of S2[1] and the detector B controls that no photons from S2 pass through the aperture F2. Being everything like this, the detector A is not influenced by the lighting conditions of S2, since no photons from it arrive there. However, it was experimentally verified that the detector A output is lower when both sources are on with respect to when only S1 is on. This difference has been ascribed to the pilot waves of the photons emitted by S2 that went through the aperture F2, reached the photons of S1 and together with their pilot waves, steered some of the photons away from the detector A.

Having these results in mind, it would be interesting and certainly easily feasible to apply the same logic to a double-slit experiment which would be modified by adding to it a further part that would play here the same role carried out in that experiment by the source S2, the detector C, the aperture F2 and the detector B. The schematic layout of the experiment would look as in Fig.2. The laser, the double slit and the photographic screen (or CCD, CMOS) are as in every double-slit experiment. The new part comes from the apparatus in Fig.1. There is a LED which is always turned on that can be screened or not. The LED is in front of a photodiode (or phototransistor or CCD) which collects the photons. The aperture[2] F2 is crucial because it is the passage through which the pilot waves of the photons of the LED may propagate and so reach the photons coming from the laser and pilot them to form different fringes. As to the type and amount of variation of the fringes nothing can be said so far. The phenomenological theory at the base of the experiments described in [4-8], and hence at the base of this experiment, predicts only the existence of the searched effect and gives indications as to the apparatus and its dimensions. One may imagine that the fringes may shift, get broader or thinner, but I will return to this issue at the end of this paragraph. The array of photodiodes are there to control the darkness of the area in front of the aperture F2, or in other words, that no photons from the LED go through the aperture.

---

[1] It makes their wave-function collapse from a quantum mechanical point of view.

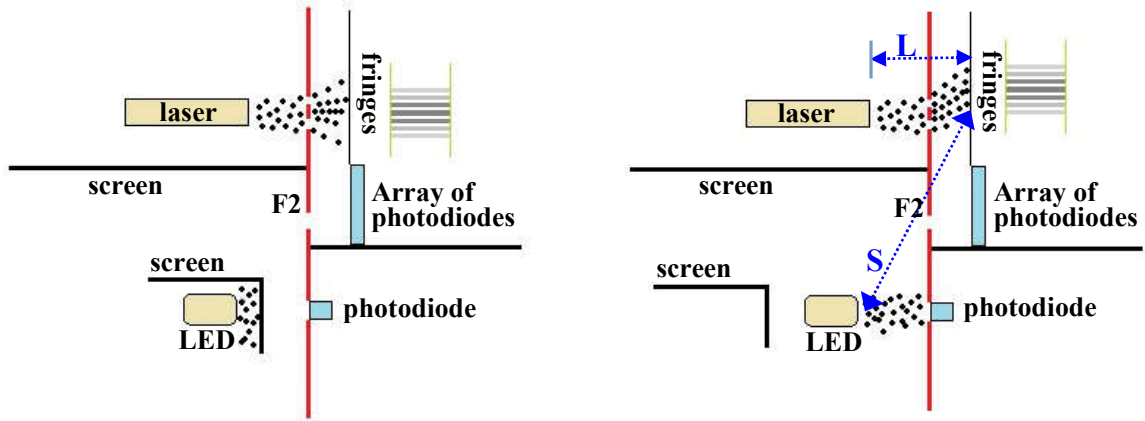

Fig2. The set-up of the double-slit experiment modified according to the apparatus in figure 1. On the left side the photons of the LED are screened along with their pilot waves. On the right side the photons of the LED are free to propagate and the same is for their pilot waves which go through the aperture F2 and modify the trajectories of the photons that form the the fringes. Hence their distribution and width should change.

I will not go into a detailed description of how the experiment has to be carried out, but I will provide only the crucial points that cannot and must not be neglected lest one fail the detection of the effect of the pilot waves. Of course all the actual measurements have to be preceded by an accurate campaign to determine what is called blank, in other words, the conditions above which the effects of the pilot waves will or should be detectable. This may sound a commonplace to the experts but I want to warn the reader and the prospective attempter, with whom I may have the pleasure to collaborate, that these effects have not been already extensively studied, conversely this attempt would be the first in the world, to the best of the author's knowledge. The only clues available are those contained in the references [4-8] and in those contained in them. As already said, in the next paragraph some theoretical explanations will be provided of what is being said here. The blank in our case is the interference pattern, the LED being screened, of which it may be advisable to collect a good amount of examples all obtained in the same conditions. It may be necessary to study with care the position, the width of the fringes and possibly the amount of photons contained in each of them. Once this is done, one may remove the screen from around the LED and, keeping the other conditions unchanged with respect to the blank, one may begin to record the interference pattern for the same amount of time. Fig.2 contains two more details: the distances L and S. These dimensions are crucial and must be the same as in the experiments of Fig.1. In these experiments these distances were varied and it turned out than there exists a range for them in order to obtain positive measurements of the effect. The range is the following: for S between 8.80 and 9.30 cm (it was decided to remain around 10 cm); for L between 1 and 4 cm [4][3]. A further information that can be drawn from [4-8] is that, if one looks at the effect in terms of energy exchanged between the photons and the pilot wave, this exchanged energy is low (it has to be less than 4.5 µeV in fact [9]). This suggests that the intensity of the laser beam above all (which of course may brought as low as the single photon conditions) but also that emitted from the LED should not be high[4], on the contrary a preliminary study should be conducted to decide the most suitable low intensities to exalt the already subtle effect.

---

[2] It is not exactly a slit. It has a finite aperture not microscopical like that of a slit.

I reckon that it is necessary to stress once more the subtlety of the experiment that is being put forward and then to speak of an other crucial point. The pilot wave is involved, according to the realistic view of quantum phenomena (Einstein, de Broglie, Bohm) in all quantum effects. This has to sound like a big warning to the experimenter's ear. Nothing is known so far about it (the pilot wave), but we do know from quantum physics that the pilot wave plays a fundamental role in all the paradoxes of quantum mechanics[5]. This tells us that the less preconceived ideas one has about what one would expect or as to how things should go according to an even long experience[6], the better. For instance, although we know something about the spatial extension over which to look for the effect and although we know that the effect is much more easily detectable if the intensity of the laser is kept low (not necessarily one photon at a time), we know absolutely nothing about the temporal evolution of a pilot wave and hence of the effects it brings about. Besides, if one considers the probabilistic nature of quantum effects (in which the pilot wave is always involved as already said), one understands better the difficulty to face in dealing with this time dependence which might hide the searched effects if not suitably considered. In [4-7] the authors mention the existence of a temporal procedure in carrying out the experiment, which consisted in the time interval to wait before starting the sampling of the signal once the LEDs had been turned on and in the duration of the sampling. In this experiment with a laser and a LED, it will be also necessary to devise a temporal procedure in order to consider the interaction of the pilot waves of the photons emitted by the LED with the pilot waves of the photons of the laser and these photons themselves. Since we cannot be sure that this interaction has always the same strength and the same sign, the purpose of the time procedure is to let the effects of this interaction be systematically visible above the background, that is the fringes may possess a visible variation (either a shift or a broadening or a shrinking or whatever it may be). The concrete actions to devise the time procedure will be those involving the turning on of the source (LED and laser), the regulation of the intensity of the laser beam and the timing of emission of the bunches of photons (above all if one works near the single photon condition).

---

[3] The reason for these distances is clearly explained in [4] but some hints will be provided later on in the paper.

[4] Far from being a phenomenon that is exalted at high energy, the pilot wave is something whose effects are better manifested at low energy. This reason is related to the connection existing between the pilot wave and the violation of the Local Lorentz Invariance. In the next paragraph something will be said about this but for a exhaustive explanation one has to refer to [4-9].

[5] The paradoxes are all connected to the wave function which, according to the realistic view of quantum mechanics (bohmian mechanics), is the pilot wave.

[6] One may be misled for instance by the name "wave" and hence expect a wave-like effect with some kind of periodicity.

One more remark has to be added here. This experiment might seem quite similar to that reported in [4] and one might wonder what new information may be achieved by it. Despite the soundness of the results of the previous experiments [4-8], which have been also corroborated by other quite different experiments performed by other teams [10,11], they were somehow rather influenced by the time procedure. On the contrary, one of the targets of this experiment is the attempt to achieve the detection of the effect 'independently' of the time procedure. This possibility is contained in the interference pattern which provides an integral (in space and time) recording of the phenomenon (either on a photographic plate or on a electronic image sensor), rather than a punctual visualisation (on a photodiode or phototransistor) in time of the intensity of the signal. Although it may sound as preconception and hence, as recommended above, should be banned from our thought, one may imagine that the pilot waves of the LED's photons act on every single photon (if working in the single photon limit) or on every small bunch of photons steering it/them away from where it/they would impinge, were the pilot waves not present. In this sense, even if the pilot waves' action were not constant in space and time (periodical or even random) the photo of the spots at the end of the recording or also the photos collected during the recording would contain anyway all the arrivals of the photons. In this sense, because of the asymmetry in the position of the LED with respect to the centre of the unaffected interference pattern, one may imagine that the effect on the initially symmetric trajectories of the photons (Bohmian mechanics), that bring about a symmetric interference pattern, may turn out to be also asymmetric and then produce some kind of asymmetry of the pattern. The second reason to perform this experiment is to somehow corroborate what was hypothesised in the previous one [6,7]. The decrease of the intensity of the photon beam of the detector A, when both sources S1 and S2 are on (see Fig.1) with respect to when only S1 is on, was suggested to be similar to a variation of the brightness of a fringe of an interference pattern, although the experimental set-up was designed to give evidence of the photon as a particle, not as a wave. The third reason lies in the necessity to find a more accurate experimental set-up, like those used in the double-slit experiments, in order to carry on the study of this fundamental feature of physics with higher and higher accuracy and precision.

**The Pilot Wave and the geometrical violation of Local Lorentz Invariance**

The purpose of this paragraph is only to sketch some of the theoretical ideas at the base of this proposal. Hence, far from being a thorough description, it will only contain a series of consecutive statements which will start from the concept of Local Lorentz Invariance (LLI) and its geometrical violation and end up to that of pilot wave, hence providing a logical thread that links quantum mechanics to the deformed local geometry of Space-Time. For a comprehensive treatment of the ideas mentioned here please refer to the references [4,8,9].

LLI is part of the Einstein Equivalence Principle of General Relativity [12,13]. It states that a free falling reference frame in a gravitational field is inertial and hence that the non gravitational laws of physics can be expressed in it in the language of Special Relativity. If Special Relativity is valid the local Space-Time is minkowskian, i.e. flat and rigid and the maximal causal speed, that coincides with the speed of light in vacuum, is invariant [9,12,13,14]. This is true if LLI is valid.

If, conversely, LLI is violated these conditions cease to be valid [9]. There can be superluminality and moreover the local Space-Time cease to be flat and rigid [9] and in general these two conditions have to be imagined to appear in the same context. In other words one is entitled (under certain conditions [9,12,13]) to think that where superluminal phenomena show up there the Space-Time is non more flat, but not curved in the sense of Riemann, rather somehow dynamically deformed [9]. For example, in an experiment carried out with two disaligned horn antennas [10] it was measured superluminal propagation of the microwave signal. In the same context one may think that the Space-Time is deformed. Two questions may rise here: can this deformation produce some kind of effects? Can these effects be measured? [4]. In order to answer these questions, the experiment of Fig.1 was conceived [4]. The conjecture at the base of it, is that the deformed Space-Time can affect the propagation of photons. This statement begins already to hint at the concept of pilot wave. As clearly described in [4] the experiment was designed on the basis of a suggestion by de Broglie and Andrade y Silva to detect the pilot wave [3] but the disposition of the sources and the detectors and their reciprocal distances were exactly the same as in the experiment of the two horn antennas where superluminality had shown up and hence where the deformed Space-Time was hypothesised to be present [4]. Thus it is here established the connection between the concept of pilot wave and that more fundamental of deformed Space-Time. From this connection it is also easily made the step to the next level where quantum mechanics and its paradoxes (all due to the lack of locality and lack of causality in the wave function description) are linked to and possibly explained by the existence of a deformed Space-Time and hence to the violation of LLI. The superluminality would cure the lack of Lorentz causality[7] by allowing a limited but unbounded maximal causal velocity, the deformed Space-Time (pilot wave) and its effects would cure the lack of the nexus of causality (i.e. the lack of the means of connection) as e.g. between entangled particles.

---

[7] With Lorentz causality it is meant the causality established with the maximal causal velocity numerically equal to the speed of light in vacuum.

## Conclusions

The experiment proposed in this paper is aimed at detecting the effects of the old debated pilot wave once detached (at least spatially) from the photons. The Pilot wave is at the heart of quantum mechanics. It is just how the realists, de Broglie, Einstein, Bohm called the mere probabilistic wave function of the conventionalists. The possibility to detect its effects means to establish once for all its reality and moreover to set the first milestone of the track that will lead to the profound realistic comprehension of quantum phenomena.

Establishing the reality of the pilot wave would indicate as a matter of fact that the quantum entities (photons, electrons and so on) are in fact real and more complex things than a mere mathematical or conceptual convention that shows up as particle or wave according to the experimental set-up.

Without depriving Quantum Mechanics of its great and so far unreachable predicting power, the capability to detect the effects of what has been called for so long 'pilot wave', (i.e. something that steers the motion of a particle, which sounds pretty much like a deformed Space-Time [9]), will set the starting point of a deep revision under three points of view: the theoretical concepts of the micro physics, including that of the geometry of the local Space-Time; the experimental capabilities to design experiments to look into a real micro world; the analytical capabilities to comprehend more thoroughly the results of the experiments, up to now reckoned to be affected by indeterminacy and complementarity.

Despite the experimental subtleties and the accuracy, mentioned above, necessary to carry out this experiment, the endeavour towards this goal is not at all unreasonable and infeasible. In the last decades, thanks to the incredible improvements in the single-photon and non-linear optics technology, it has been reached the suitability to embark on this attempt. Under this point of view, Julius Robert Oppenheimer once said: "*It is a profound and necessary truth that the deep things in science are not found because they are useful; they are found because it was possible to find them.*"

## Acknowledgements
It is with the great pleasure that I thank professor Gianni Albertini for his remarks and suggestions.